\documentclass{appolb}
\usepackage{graphicx}
\usepackage{wrapfig}
\usepackage{amsmath}
\usepackage{lineno}
\usepackage[hidelinks]{hyperref}
\hypersetup{
  colorlinks=true,
  citecolor=blue,
  urlcolor=blue,
  linkcolor=blue
}
\begin{document}
\title{Recent ALICE results from light-ion collision systems%
  \thanks{Presented at the 32$^{\mathrm{nd}}$ Cracow Epiphany Conference on the recent results from Heavy Ion Physics, Krak\'ow, Poland, 12--16 January, 2026.}% Krak$\acute{o}$w
}
\author{Abhi Modak
  \address{on behalf of the ALICE Collaboration}
  \address{INFN, Sezione di Trieste, Trieste, Italy}
}
\headauthor{Abhi Modak}
\headtitle{}
\maketitle
%\linenumbers
\begin{abstract}
  This article presents recent measurements by the ALICE Collaboration in proton--oxygen (pO), oxygen--oxygen (OO), and neon--neon (Ne--Ne) collisions delivered by the LHC in July 2025. Measurements of the primary charged-particle pseudorapidity density and the elliptic and triangular flow coefficients of charged particles are reported. Experimental evidence of the suppression of neutral pion yields in OO collisions relative to the proton--proton baseline is also discussed. Comparisons of these new data with theoretical models provide key input to understand particle production, collective phenomena, and parton energy loss in small collision systems.
\end{abstract}

\section{Introduction}
Recent results from the LHC experiments in high-multiplicity proton--proton (pp) and proton--lead (p--Pb) collisions have shown features such as collective flow and strangeness enhancement, which are usually attributed to the formation of the quark-gluon plasma (QGP) in heavy-ion collisions (see e.g.~\cite{ALICE:2022wpn} and references therein). However, measurements in these small collision systems have not shown clear evidence of jet quenching (arising from the in-medium interactions of energetic partons in dense matter) within current experimental uncertainties. Since jet quenching is a expected consequence of the QGP formation, this tension between physics pictures from measurements of soft and hard (high-$Q^2$) observables in small systems has become a key question in the LHC heavy-ion programme.

In this context, collisions of light ions ($^{16}$O and $^{20}$Ne) at the LHC provide a unique opportunity to explore the effects seen in high-multiplicity pp and p--Pb collisions, with a system that has a similarly small number of participating nucleons and similar final-state multiplicity but with larger geometrical transverse overlap, thereby enhancing the jet-quenching effects, which depend on the in-medium path length. In addition to searching for jet quenching effects, these OO and Ne--Ne collisions offer the possibility to investigate collective flow effects and to study particle production in a multiplicity range that bridges pp and p--Pb on the low side, and Pb--Pb and Xe--Xe on the high side.

In this article, we report three key measurements performed by the ALICE experiment in light-ion collision systems: (a) the primary charged-particle pseudorapidity density ($\mathrm{d}N_{\mathrm{ch}}/\mathrm{d}\eta$); (b) elliptic ($v_2$) and triangular flow ($v_3$) coefficients of charged particles; and (c) the suppression of neutral pion production in OO relative to pp collisions.

\section{Experimental details}
The ALICE apparatus has been upgraded during the Long Shutdown~2 of the LHC~\cite{ALICE:2023udb}. Key upgrades include a new tracking system in the central barrel~\cite{ALICE:2013nwm,Lippmann:2014lay}, a new forward trigger detector~\cite{Trzaska:2017reu}, continuous readout~\cite{Antonioli:2013ppp}, and a novel online-offline software framework~\cite{ALICE:o2} to handle both data reconstruction and analysis. The main sub-detectors used in these analyses are the Inner Tracking System (ITS)~\cite{ALICE:2013nwm}, the Time Projection Chamber (TPC)~\cite{Lippmann:2014lay}, the Fast Interaction Trigger (FIT)~\cite{Trzaska:2017reu}, and the Electromagnetic Calorimeter (EMCal)~\cite{ALICE:2022qhn}.\,The ITS and the TPC detectors are used to reconstruct charged particles at midrapidity for measuring $\mathrm{d}N_{\mathrm{ch}}/\mathrm{d}\eta,v_2$,\,and\,$v_3$.\,The EMCal is used to reconstruct photons from $\pi^0$ decays. The relevant FIT components are the FT0A and FT0C, which provide precise timing for continuous readout and are also used for centrality estimation, event selection, and collision time determination. Minimum-bias events are selected when both FT0A and FT0C arrays detect at least one hit in the nominal collision time window. Collisions are removed based on their proximity in time to the ITS readout frame border and the time frame border. Events with a longitudinal position within 10\,cm of the nominal interaction point are selected. Additionally, events where the estimate of vertex position from FIT is more than 1\,cm away from the tracked vertex position are ignored due to lower overall quality. The selected event sample is then classified into centrality classes using the raw amplitudes of signals in the FT0C detector, which are proportional to the number of charged particles detected by the FT0C. 

\section{Charged-particle pseudorapidity density at midrapidity}
\label{dndeta}
ALICE has measured the pseudorapidity density of primary charged particles, $\mathrm{d}N_{\mathrm{ch}}/\mathrm{d}\eta$, in pO, OO, and Ne--Ne collisions as a function of pseudorapidity and studied its dependence on the event centrality from central (0--5\%) to peripheral (60--90\%) collisions~\cite{M-Urioni:2025IS}. In this analysis, tracks are categorised into three exclusive classes: global tracks (built from ITS and TPC contributions), ITS-only tracks, and TPC-only tracks. TPC-only tracks are excluded because their pointing resolution is low, as the innermost TPC radius is quite far from the collision vertex, and they cannot be reliably assigned to a collision. The selected track sample contains global tracks and ITS-only tracks that pass track quality cuts and satisfy a criterion on the distance of closest approach to ensure they originate from primary particles~\cite{ALICE:2025cjn}. The raw $\mathrm{d}N_{\mathrm{ch}}/\mathrm{d}\eta$ is then corrected for detector acceptance and efficiency using simulations with PYTHIA 8.3/Angantyr~\cite{Bierlich:2018xfw}, following the method described in Ref.~\cite{ALICE:2025cjn}. Systematic uncertainties from various sources (extrapolation to zero $p_{\rm T}$, particle composition, variations in detector acceptance, uncertainty on the centrality determination, variations in track selection) are evaluated and then added in quadrature to obtain the total systematic uncertainty, which is 3.5\% (10\%) for the central (peripheral) collisions for all the colliding systems.

\begin{wrapfigure}{r}{0.45\textwidth}
  \vspace*{-1.1cm}
  \begin{center}
    \includegraphics[scale=0.3]{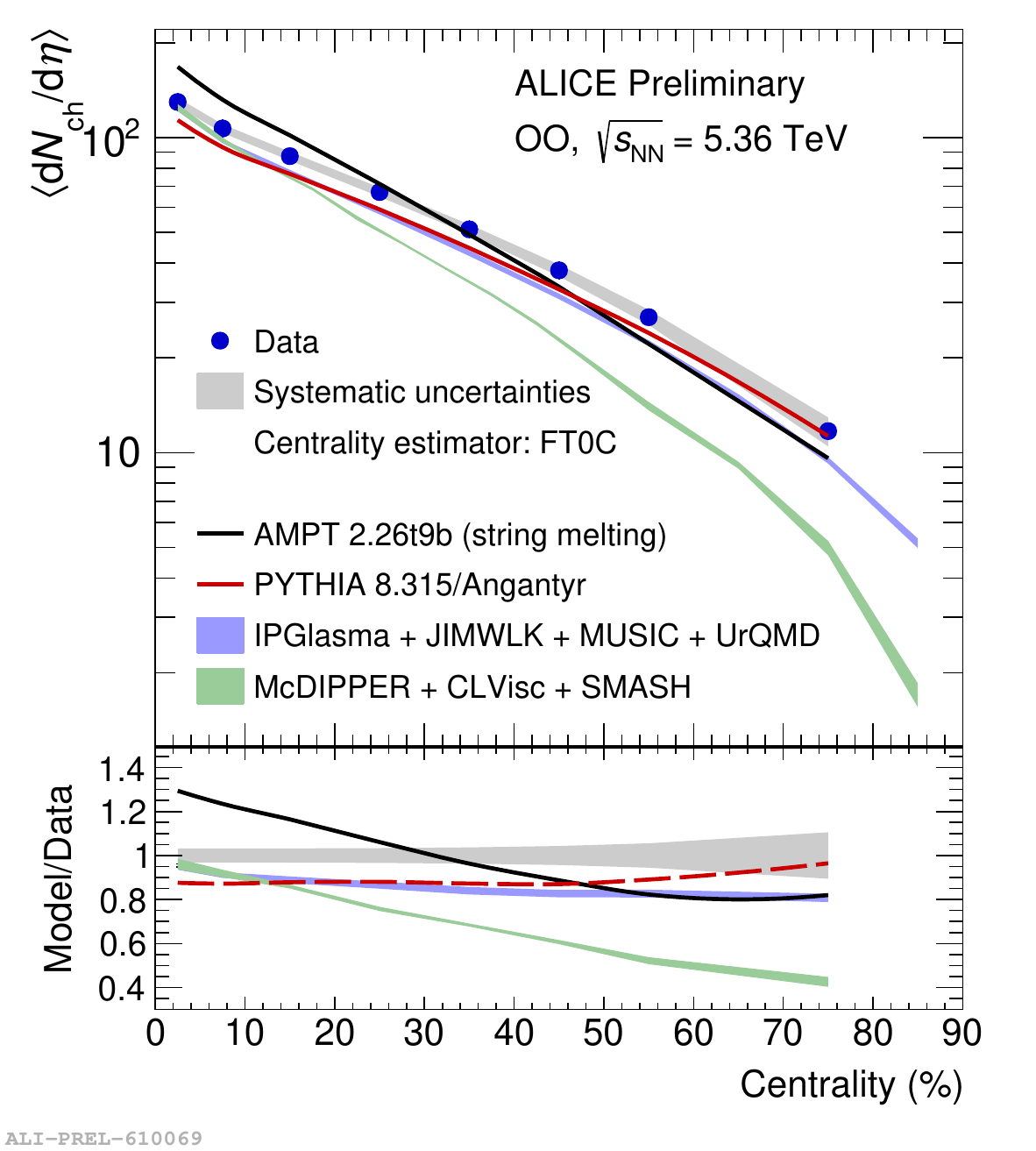}
  \end{center}
  \vspace*{-0.4cm}
  \caption{Dependence of $\langle \mathrm{d}N_{\mathrm{ch}}/\mathrm{d}\eta \rangle$ on the collision centrality in OO collisions at $\sqrt{s_{\mathrm{NN}}}=5.36$ TeV. Grey bands show the systematic uncertainties. Statistical uncertainties are within the marker size. Predictions from various theoretical models are superimposed.}
  \label{avgdndeta_OO}
\end{wrapfigure}

The resulting charged-particle pseudorapidity density, averaged over \linebreak $|\eta|<0.5$ and denoted by $\langle \mathrm{d}N_{\mathrm{ch}}/\mathrm{d}\eta \rangle$, is presented in Fig.~\ref{avgdndeta_OO} as a function of centrality in OO collisions at $\sqrt{s_{\mathrm{NN}}}=5.36$\,TeV. The measured $\langle \mathrm{d}N_{\mathrm{ch}}/\mathrm{d}\eta \rangle$ are $129.7\pm 4.1$\,(syst) and $11.7\pm1.2$\,(syst) for the 0--5\% and 60--90\% centrality classes, respectively. The statistical uncertainties are negligible. The data are compared to predictions from the PYTHIA 8.3/Angantyr~\cite{Bierlich:2018xfw} and AMPT~\cite{Lin:2004en} event generators and to hydrodynamic calculations using IPGlasma~\cite{Schenke:IPglasma1} and McDIPPER~\cite{Garcia-Montero:MCdipper1} models. Both IPGlasma and McDIPPER models provide a reasonable description of the central OO data within the uncertainties. However, McDIPPER fails to reproduce both the shape and the magnitude of the centrality dependence of $\langle \mathrm{d}N_{\mathrm{ch}}/\mathrm{d}\eta \rangle$ from mid-central to peripheral collisions. AMPT also shows a trend inconsistent with the data whereas PYTHIA 8.3/Angantyr qualitatively captures the overall trend and agrees with the measured value in peripheral collisions. Similar results for pO and Ne--Ne collisions (not shown here) can be found in Ref.~\cite{M-Urioni:2025IS}.

\begin{figure}[h!]
  \centering
  \includegraphics[scale=0.3]{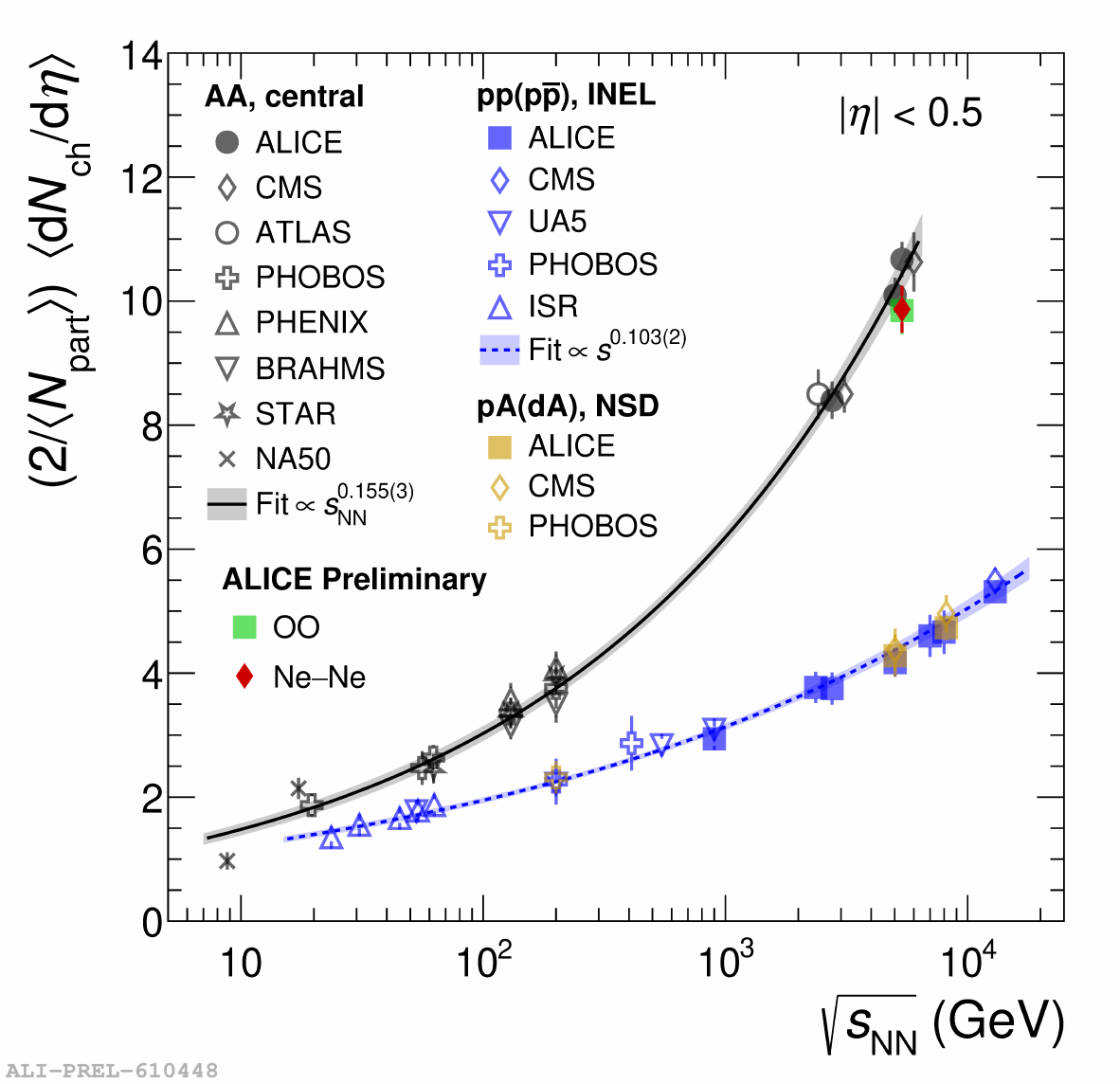}
  \includegraphics[scale=0.3]{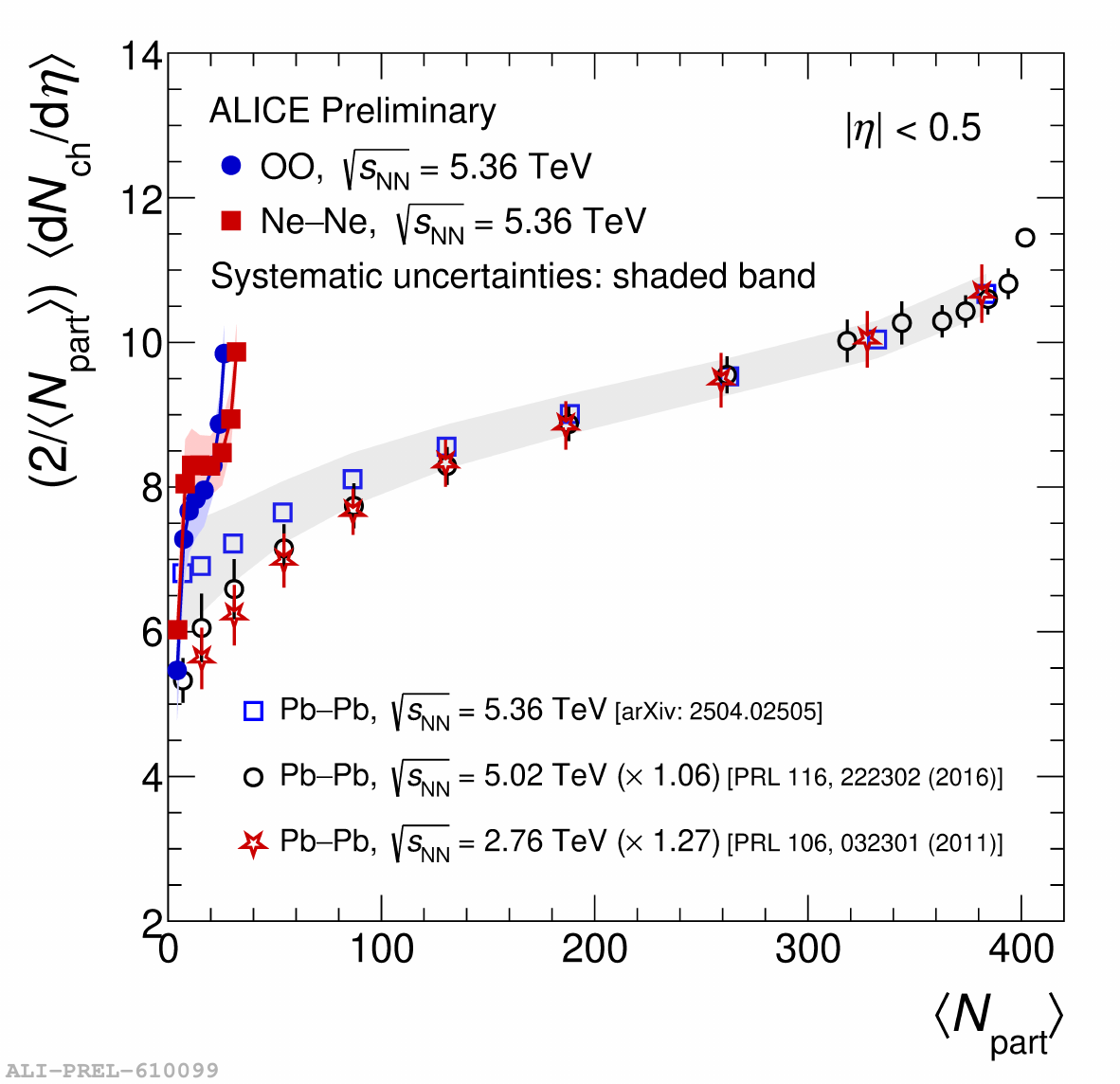}
  \caption{(Left) Values of ($2/\langle N_{\mathrm{part}} \rangle)\langle \mathrm{d}N_{\mathrm{ch}}/\mathrm{d}\eta \rangle$ for the 5\% central OO and Ne--Ne collisions compared to previous measurements in pp(p$\bar{p}$), pA(dA), and central heavy-ion collisions as a function of $\sqrt{s_{\mathrm{NN}}}$. (Right) Dependence of ($2/\langle N_{\mathrm{part}} \rangle)\langle \mathrm{d}N_\mathrm{ch}/\mathrm{d}\eta \rangle$ on the $\langle N_{\mathrm{part}} \rangle$ for OO and Ne--Ne collisions at $\sqrt{s_{\mathrm{NN}}}=5.36$ TeV. ALICE data from Pb--Pb collisions at the LHC energies are shown for comparison.}
  \label{avgdndeta_OOcent}
\end{figure}

To compare the bulk particle production in different collision systems that collided at different energies, the $\langle \mathrm{d}N_{\mathrm{ch}}/\mathrm{d}\eta \rangle$ is normalised by the average number of participating nucleon pairs, $\langle N_{\mathrm{part}} \rangle/2$, determined for each centrality class using the Glauber model calculation. In Fig.~\ref{avgdndeta_OOcent} (left), values of the normalised multiplicity, ($2/\langle N_{\mathrm{part}} \rangle)\langle \mathrm{d}N_{\mathrm{ch}}/\mathrm{d}\eta \rangle$, in central (0--5\%) OO and Ne--Ne collisions are compared to the data from pp(p$\bar{p}$), pA(dA), and central heavy-ion collisions at lower or similar centre-of-mass energies (see~\cite{ALICE:2025cjn} and references therein). The OO and Ne--Ne results are in agreement within uncertainties with the trend established from previous heavy-ion measurements, which shows a stronger rise as a function of $\sqrt{s_{\mathrm{NN}}}$ than for pp and pA collisions. Figure~\ref{avgdndeta_OOcent} (right) shows the measured ($2/\langle N_{\mathrm{part}} \rangle)\langle \mathrm{d}N_{\mathrm{ch}}/\mathrm{d}\eta \rangle$ as a function of $\langle N_{\mathrm{part}} \rangle$ in OO (solid blue) and Ne--Ne (solid red) collisions. A pronounced centrality dependence is observed, with ($2/\langle N_{\mathrm{part}} \rangle)\langle \mathrm{d}N_{\mathrm{ch}}/\mathrm{d}\eta \rangle$ decreasing by a factor of $\simeq$1.7 from central (large $\langle N_{\mathrm{part}} \rangle$) to peripheral (small $\langle N_{\mathrm{part}} \rangle$) collisions. The results are compared with previous ALICE measurements in Pb--Pb collisions, which show that the normalised multiplicity in OO and Ne--Ne collisions increases more steeply than in heavy-ion data at the same $\langle N_{\mathrm{part}} \rangle$ values.

\section{Anisotropic flow in light-ion collisions}
One of the experimental observables that is sensitive to the properties of the QGP medium is the anisotropic flow, which arises from the transfer of initial spatial anisotropy to momentum anisotropy in the final state via pressure gradients within the created medium. It is characterised by the Fourier coefficients $v_n$ of the azimuthal particle distribution,

\begin{equation*}
  \frac{dN}{d\varphi} \propto 1 + 2 \sum_{n=1}^{\infty} v_n cos[n(\varphi - \Psi_n)],
\end{equation*}
where $\varphi$ is the azimuthal angle of the final-state produced particle, $\Psi_n$ is the $n^{\rm th}$-order symmetry plane, and $v_n$ are the flow coefficients. The elliptic ($v_2$), and triangular ($v_3$) flow coefficients are measured in OO and Ne--Ne collisions for unidentified charged particles reconstructed by ITS and TPC detectors~\cite{ALICE:2025luc}. Similar track selection criteria are applied as described in Sec.~\ref{dndeta}. The flow measurements are obtained from two- and multiparticle cumulants using the generic framework~\cite{Bilandzic:2013kga}, which corrects for the effects due to detector acceptance and track reconstruction efficiency. A pseudorapidity separation of $|\Delta\eta|>1.4$ is applied in the two-particle correlation measurements~\cite{Zhou:2015iba} which suppresses correlations from jets and resonance decays. Systematic uncertainties are evaluated using relative differences by varying the event and track selection criteria and are added to the estimates of the remaining nonflow effects. The resulting systematic uncertainties up to 6.5\% are assigned to $v_n$\{2\} and $v_n$\{4\} for both OO and Ne--Ne collisions.

\begin{figure}[h!]
  \begin{minipage}[c]{0.5\textwidth}
    \includegraphics[scale=0.45]{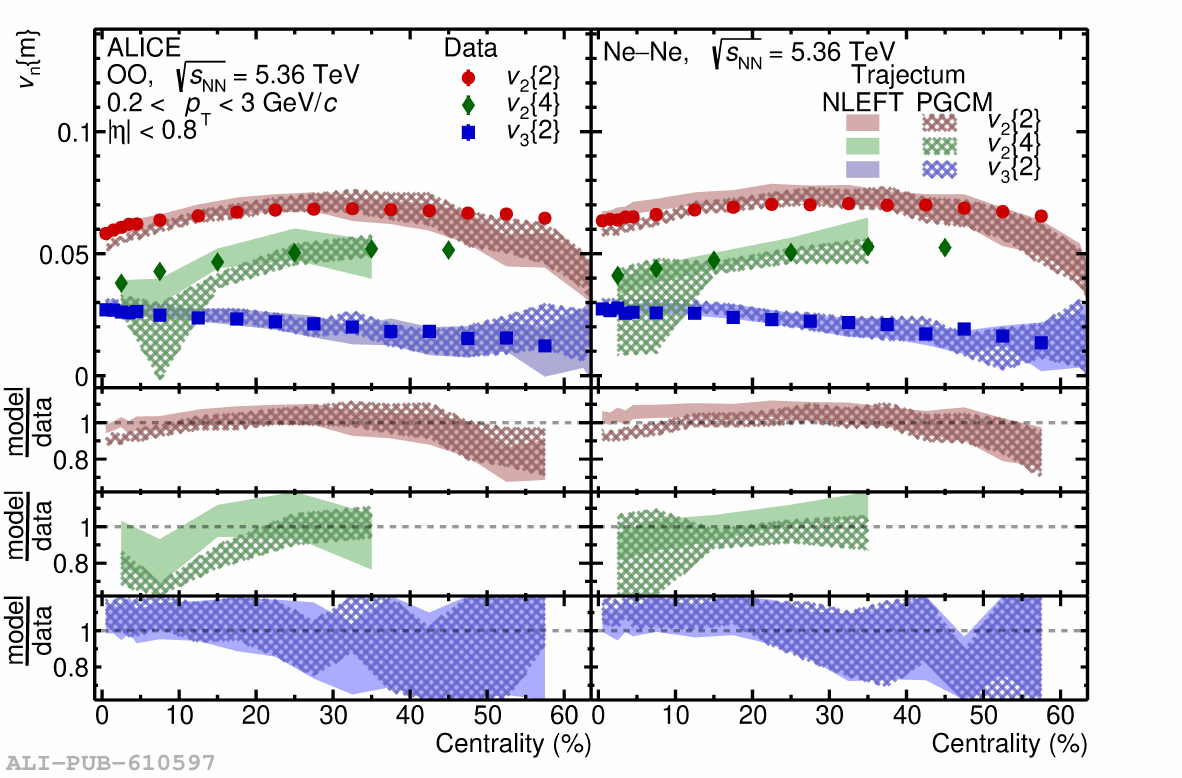}
  \end{minipage}\hfill
  \begin{minipage}[c]{0.32\textwidth}
   \caption{Centrality dependence of $v_2$\{2\}, $v_3$\{2\}, and $v_2$\{4\} for charged particles with $0.2<p_{\rm T}<3$\,GeV/$c$ and $|\eta|<0.8$ in OO (left) and Ne--Ne (right) collisions at $\sqrt{s_{\mathrm{NN}}}=5.36$\,TeV. The measurements are compared with Trajectum calculations with NLEFT and PGCM inputs.}
   \label{flow_cent}
   \end{minipage}
\end{figure}

In Fig.~\ref{flow_cent}, the charged particle $v_2$\{2\}, $v_3$\{2\}, and $v_2$\{4\} are presented as a function of centrality in OO (left) and Ne--Ne (right) collisions. In both systems, $v_2$\{2\} exhibits a weak centrality dependence with a magnitude that slightly increases from central to mid-central collisions and then decreases in peripheral collisions, following the trend predicted by the initial state eccentricity calculation~\cite{Loizides:2025ule}. The coefficient $v_2$\{4\} is non-zero and systematically smaller than the $v_2$\{2\} values. The non-zero $v_2$\{4\} values indicate the collective nature of the measured anisotropy and confirm the presence of anisotropic flow in OO and Ne--Ne collisions. Unlike $v_2$, the measured $v_3$\{2\} increases from peripheral to central collisions, opposite to the trend predicted by the initial state triangular eccentricity calculation~\cite{Loizides:2025ule}. The magnitude of $v_3$\{2\} is similar to that observed in pp and p--Pb collisions~\cite{ALICE:2019zfl}. This suggests that initial state fluctuations, which lead to $v_3$\{2\}, may be similar between small and light-ion systems at the same multiplicity. In Fig.~\ref{flow_cent}, the measurements are also compared with hydrodynamic simulations from the Trajectum framework~\cite{Nijs:2021clz}, which incorporate nuclear structure inputs for $^{16}$O and $^{20}$Ne derived from the Nuclear Lattice Effective Field Theory (NLEFT)~\cite{Lee:2025req} and the ab initio Projected Generator Coordinate Method (PGCM)~\cite{Frosini:2021fjf}. NLEFT-based calculations reproduce both the trend and magnitudes of $v_2$\{2\}, $v_3$\{2\}, and $v_2$\{4\} in the considered centrality range, while PGCM-based results show slightly worse agreement in central collisions, though the deviation from the data remains within a few percent. The consistency between the presented $v_2$\{2\}, $v_3$\{2\}, and $v_2$\{4\} measurements and hydrodynamic model predictions supports the emergence of collective behaviour in light-ion collisions.

%\vspace{-0.6cm}
\section{Observation of parton energy loss}
This section presents the measurements of invariant yields of neutral pions at midrapidity ($|y|<0.8$) in the $p_{\rm T}$ range $1.2<p_{\rm T}<20$\,GeV/$c$ in pp and OO collisions at the same centre-of-mass energy. The modification of the $\pi^0$ yields for different $p_{\rm T}$ intervals in OO collisions with respect to pp collisions can be quantified by the nuclear modification factor

\begin{equation*}
 R_{\rm OO} (p_{\rm T}) = \frac{\mathrm{d}^2\sigma / \mathrm{d}p_{\rm T}\mathrm{d}y|_{\mathrm{OO}}}{\langle T_{\mathrm{OO}} \rangle \times \mathrm{d}^2\sigma / \mathrm{d}p_{\rm T}\mathrm{d}y|_{\mathrm{pp}}}
\end{equation*}
where the nuclear overlap function $\langle T_{\mathrm{OO}} \rangle$ is related to the average number of inelastic nucleon--nucleon collisions as $\langle T_{\mathrm{OO}} \rangle = \langle N_{\mathrm{coll}} \rangle / \sigma_{\mathrm{inel}}^{\mathrm{pp}}$. The EMCal is used to reconstruct $\pi^0$ via the two-photon decay channel, $\pi^0 \rightarrow \gamma\gamma$, with a branching ratio of 98.8\%, using the invariant mass technique. The resulting photon-pair distribution contains both the $\pi^0$ signal and combinatorial background from pairs not originating from the same $\pi^0$. The background is estimated with the rotational event mixing method~\cite{ALICE:2022qhn}. After background subtraction, the $\pi^0$ yield ($N^{\pi^0}$) is obtained by integrating the counts over a $\pm3\sigma$ interval around the reconstructed $\pi^0$ peak. In order to obtain the invariant yield, the $N^{\pi^0}$ is corrected for the detector acceptance, efficiency, and contamination from secondary $\pi^0$ contributions (i.e., $\pi^0$ produced in interactions with detector material or from long-lived ($c\tau>1$\,cm) particle decays). Systematic uncertainties from various sources (photon reconstruction, $\pi^0$ signal extraction, and material budget) are evaluated. However, the uncertainty of the material budget (4.2\%) cancels in the nuclear modification factor. A conservative normalisation uncertainty of the production cross section (10\% in OO and 6\% in pp) is propagated to the $R_{\rm OO}$ measurement. 

\begin{figure}[h!]
  \centering
  \includegraphics[scale=0.29]{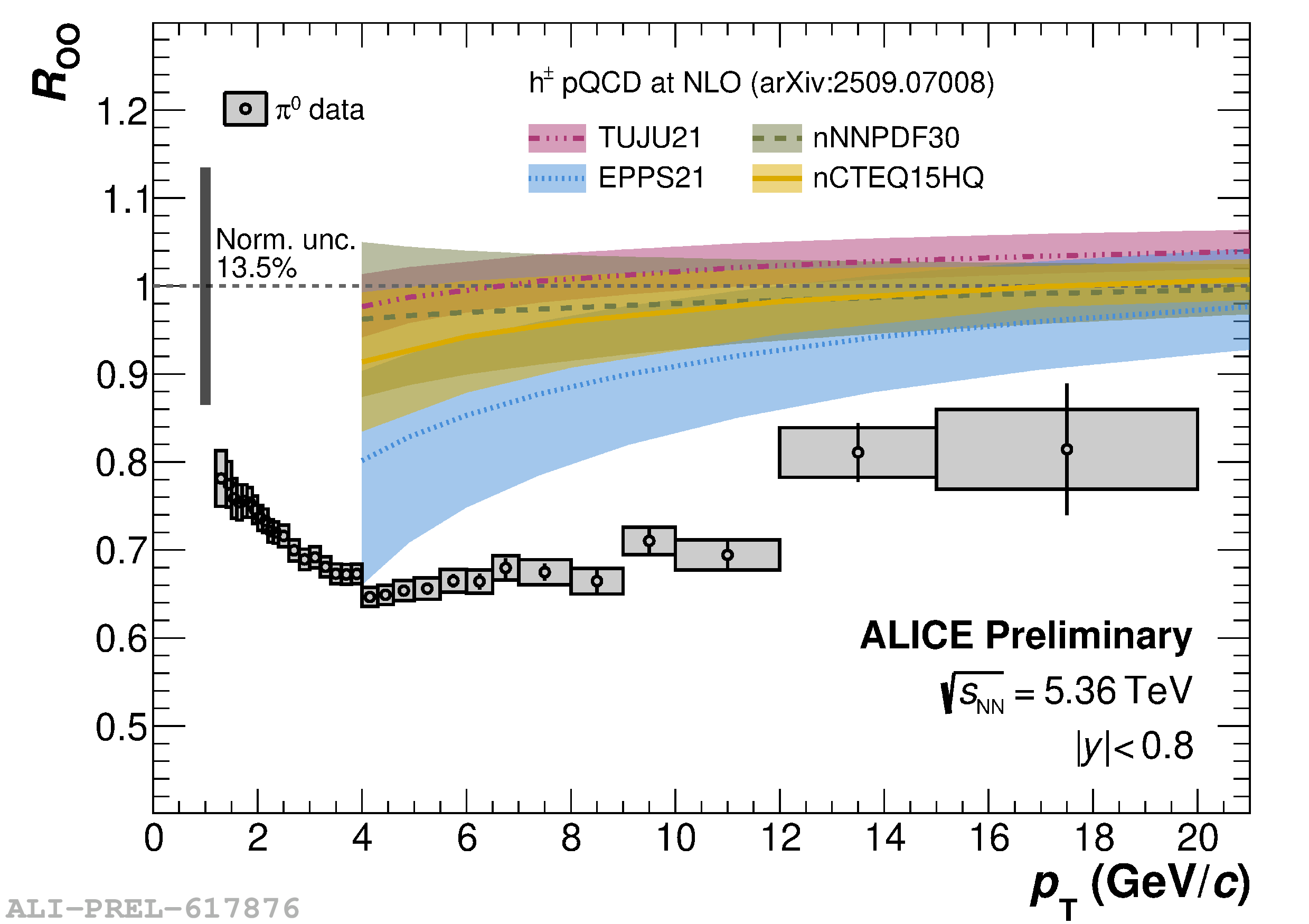}
  \includegraphics[scale=0.29]{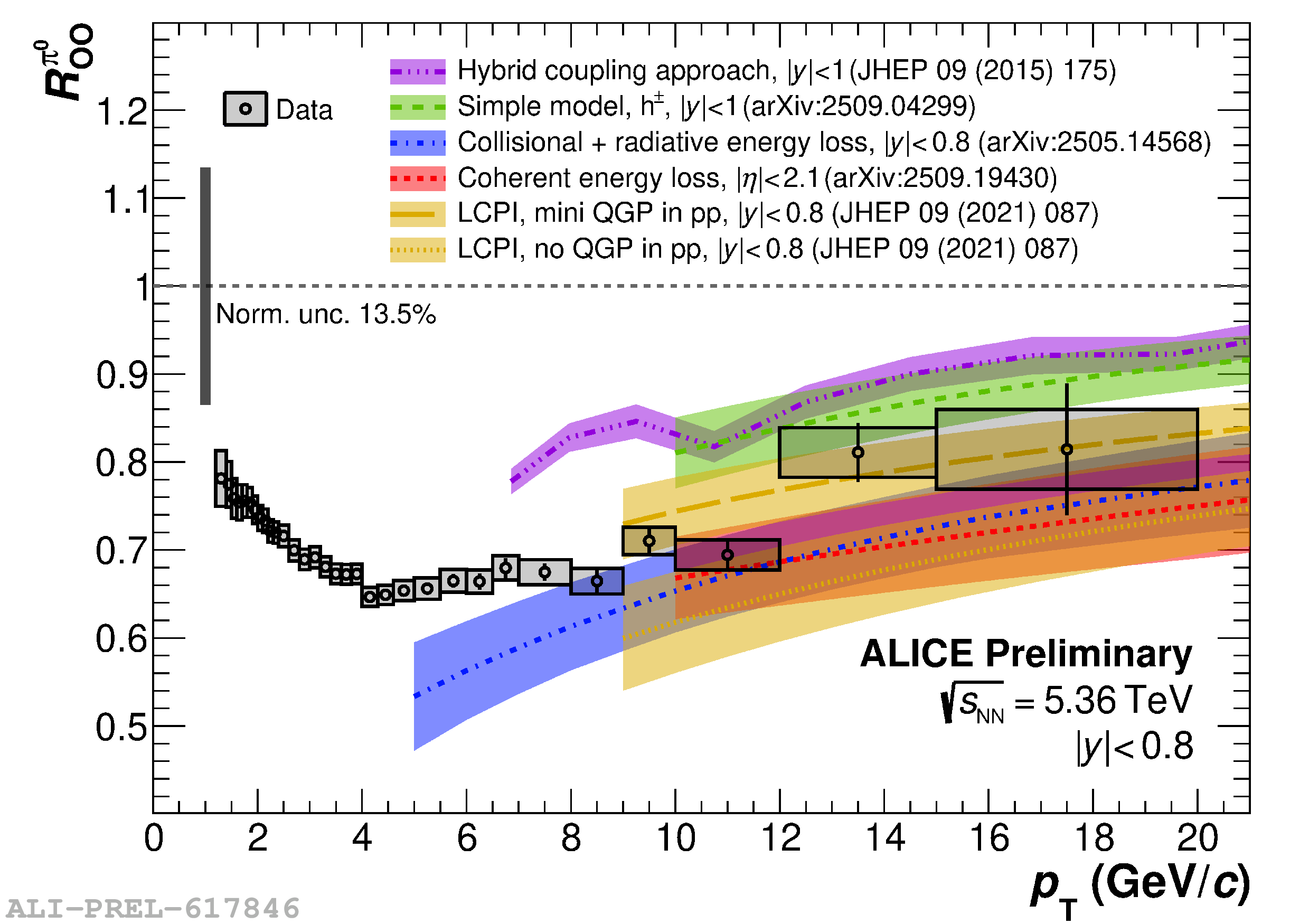}
  \caption{Measured $R_{\rm OO}$ of $\pi^0$ in comparison to various theoretical predictions. Models in the left plot are baselines that do not include parton energy loss, while models on the right include energy loss. Model references are included in the figure.}
  \label{RooFig}
\end{figure}

Figure~\ref{RooFig} presents the measured $R_{\rm OO}$ of $\pi^0$ as a function of $p_{\rm T}$ in OO collisions, which shows a clear suppression relative to unity. In the left panel, the data are compared with four next-to-leading order (NLO) pQCD calculations using nuclear parton distribution functions (nPDFs) that include cold nuclear matter (CNM) effects. Calculations based on TUJU21, nNNPDF30, and nCTEQ15HQ deviate by at most about 5\% from unity, while the EPPS21 prediction shows a larger deviation of up to 20\%. Considering the sizable nPDF uncertainties, indicated by the shaded bands, the observed suppression of $\pi^0$ production in OO collisions beyond CNM expectations reaches a significance of 2.4$\sigma$ when compared with the most suppressed prediction based on EPPS21. In the right panel, the measured $R_{\rm OO}$ is compared with models that incorporate the parton energy loss through different mechanisms. These energy-loss models use EPPS21 to include the CNM effects. Therefore, the interpretation of the model spread must account for the fact that the predicted modifications reflect both the modeled hot-medium energy loss and the assumed nPDF baseline. Overall, models capture the $p_{\rm T}$-dependent shape of the $R_{\rm OO}$ for $p_{\rm T}>10$\,GeV/$c$. Future measurements of the nuclear modification factor in pO collisions will provide experimental constraints on CNM effects in oxygen and improve the interpretation of the observed suppression.

%\vspace{-0.4cm}
\section{Summary}
In summary, new measurements from the ALICE Collaboration in pO, OO, and Ne--Ne collisions at the LHC provide important insights into particle production and medium effects in light-ion systems. The charged-particle multiplicity follows the energy-dependent trend previously observed in heavy-ion collisions, while anisotropic flow measurements are consistent with hydrodynamic predictions that include a collectively expanding medium. Furthermore, the suppression of neutral pion yields in OO collisions relative to the pp baseline suggests the presence of parton energy loss effects beyond the CNM contributions. These ALICE measurements represent an important step forward and highlight the importance of collecting light nuclei data at the LHC.

%%%%%%%% Bibliography 
\bibliographystyle{utphys}   % Remember we use title in the biblio
\bibliography{bibliography}

\end{document}